\documentclass[twocolumn]{revtex4}
\usepackage[dvips]{graphicx}
\usepackage{float, color,array}

\begin{document}

\title{Spin-lattice-coupled order in Heisenberg antiferromagnets on the pyrochlore lattice}

\author{Kazushi Aoyama and Hikaru Kawamura}

\date{\today}

\affiliation{Department of Earth and Space Science, Graduate School of Science, Osaka University, Osaka 560-0043, Japan
}

\begin{abstract}
Effects of local lattice distortions on the spin ordering are investigated for the antiferromagnetic classical Heisenberg model on the pyrochlore lattice. It is found  by Monte Carlo simulations that the spin-lattice coupling (SLC) originating from site phonons induces a first-order transition into two different types of collinear magnetic ordered states. The state realized at stronger SLC is cubic symmetric characterized by the magnetic  $(\frac{1}{2},\frac{1}{2},\frac{1}{2})$ Bragg peaks, while that at weaker SLC is tetragonal symmetric characterized by the $(1,1,0)$ ones, each accompanied by the commensurate local lattice distortions. Experimental implications to chromium spinels are discussed.
\end{abstract}

\maketitle

 In frustrated magnets, spins are often coupled to other degrees of freedom in solids. A series of spinel oxides AB$_2$O$_4$ provides typical examples of the coupling between the spin and the lattice or orbital degrees of freedom, where the magnetic ion B$^{3+}$ forms a pyrochlore lattice, a three-dimensional network of corner-sharing tetrahedra. Since the pioneering work by Yamashita and Ueda \cite{SLC_Yamashita_00}, it has been realized that the spin and the lattice often conspire to resolve frustration, giving rise to the spin-lattice-coupled ordering. In chromium spinels ACr$_2$O$_4$ (A=Zn, Cd, Hg, Mg), the orbital channel of Cr$^{3+}$ is off because of the half-filled $t_{2g}$ level, so that this system serves as a platform to investigate fundamental physics of the spin-lattice-coupled order. In this paper, bearing chromium spinels in our mind, we consider the effect of lattice distortions on the long-range spin ordering.

 Since Cr$^{3+}$ has spin-$3/2$ and a relatively weak magnetic anisotropy, the classical Heisenberg model should provide a reasonable modelling. It is theoretically established that the classical Heisenberg spins on the pyrochlore lattice with the antiferromagnetic (AF) nearest-neighbor (NN) interaction do not order at any finite temperature due to the massive degeneracy of the ground state \cite{Reimers_MC_92, Moessner-Chalker_prl, Moessner-Chalker_prb}. Weak perturbative interactions such as further-neighbor interactions would lift the degeneracy, eventually leading to the magnetic ordering, but such interactions are abundant in nature and the mechanism of the degeneracy lifting would generally depend on specific materials. % In this paper, with chromium spinels in mind, we consider the effect of the coupling of the lattice to the spin.

 In the ACr$_2$O$_4$ compounds, a common ordering feature has been observed in experiments: they undergo a first-order transition into the magnetic long-range-ordered state accompanied by a structural transition which lowers the original cubic crystal symmetry \cite{ZnCrO_Lee_00, CdCrO_Chung_05, HgCrO_Ueda_06, MgCrO_Ortega_08}. Similar magnetostructural orderings have also been observed in the ``breathing pyrochlore'' lattice consisting of an alternating array of small and large tetrahedra, LiInCr$_4$O$_4$ and LiGaCr$_4$O$_4$ \cite{BrPyro_Okamoto_13, BrPyro_Tanaka_14, BrPyro_Okamoto_15, BrPyro_Nilsen_15} (in LiInCr$_4$O$_4$, the structural transition preempts the magnetic one). In spite of the spin-lattice coupling (SLC) commonly seen in these chromium spinels, the spin-ordering patterns vary from material to material \cite{ZnCrO_Lee_08, CdCrO_Chung_05, HgCrO_Matsuda_07, MgCrO_Ortega_08, BrPyro_Nilsen_15}, and the origin of the magnetic orderings has not been well understood. In view of such an experimental situation, we theoretically investigate the SLC effect in a simple AF Heisenberg model with a {\it local\/} lattice distortion to shed light on the nature of the orderings of chromium spinels.

 Most of previous theoretical studies on the SLC in the pyrochlore antiferromagnet can be categorized into two streams: one is a phenomenological theory based on Ref.\cite{SLC_Yamashita_00} \cite{SLC_Tchernyshyov_prl_02, SLC_Tchernyshyov_prb_02} and the other is a more microscopic theory based on the so-called ``bond-phonon'' model \cite{Bond_Penc_04, Bond_Motome_06, Bond_Shannon_10}. The latter has bearing on our work taking account of the local lattice distortion. In the bond-phonon model, the lattice deformation is assumed to occur independently at each bond. It turns out that the NN classical Heisenberg model with a bond-phonon coupling yields a bond-nematic-type ordered state, without accompanying the magnetic long-range order (LRO), in contrast to the experimental result. Then, the inclusion of the ferromagnetic third-neighbor interaction turned out to lift the massive degeneracy of the NN model, leading to the collinear magnetic ordered state. In particular, in-field properties of such a model are qualitatively consistent with the experimental results on ACr$_2$O$_4$ \cite{ZnCrO_Miyata_jpsj_11, ZnCrO_Miyata_prl_11, CdCrO_Kojima_08, HgCrO_Ueda_06}.

 While independent bond-length vibrations are assumed in the bond-phonon model, in reality, a magnetic ion at each {\it site\/} vibrates implying a strong correlation among the surrounding bond lengths. A counter model of ``site phonon'', where the Einstein model was assumed for the lattice-vibration part, was considered in Refs.\cite{Site_Jia_05, Site_Bergman_06, Site_Wang_08}. The ordering properties of the site-phonon model, however, have still remained unclear, and in this paper we will determine its zero-field phase diagram for classical Heisenberg spins.

\begin{figure}[t]
\includegraphics[width=\columnwidth]{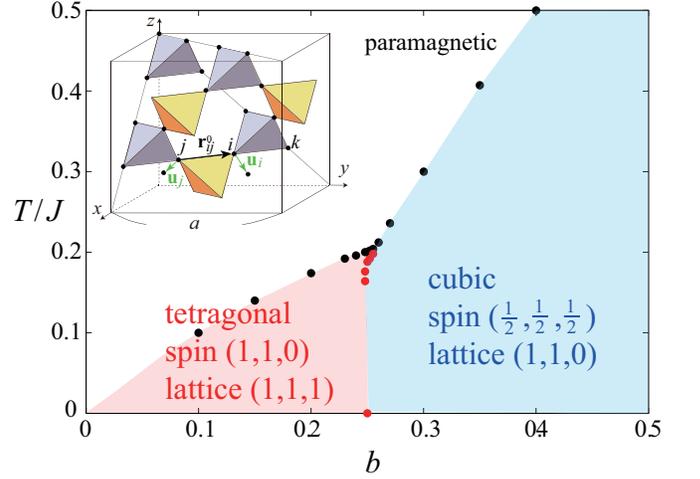}
\caption{The $b$-$T$ phase diagram of the pyrochlore antiferromagnet with local lattice distortions, where the parameter $b$ measures the strength of the spin-lattice coupling and $T$ the temperature. Two types of collinear magnetic long-range-ordered phases with the commensurate lattice distortions are realized: see the main text for details. The black dots denote first-order transition points between the paramagnetic and the ordered phases, while the red dots denote those between the ordered phases. The inset shows a cubic unit cell of the side length $a$, where the lattice distortion vector ${\bf u}_i$ is represented by the green arrow. \label{fig:b-T}}
\end{figure}

\begin{figure}[t]
\includegraphics[width=\columnwidth]{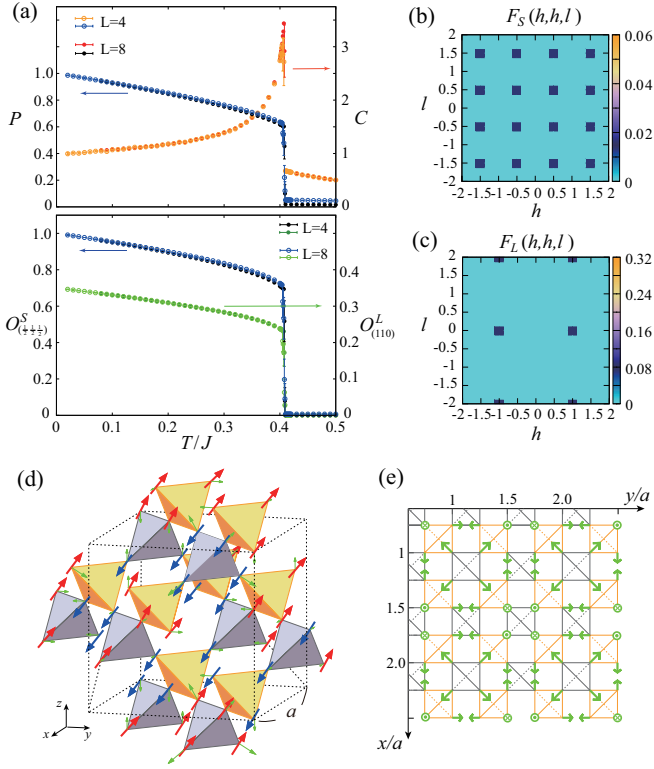}
\caption{MC results for $b=0.35$ obtained in warming runs. (a) The temperature dependences of the specific heat $C$, the spin collinearity $P$ (upper panel), and the average spin and lattice Bragg intensities $O^S_{(\frac{1}{2}\frac{1}{2}\frac{1}{2})}$ and $O^L_{(110)}$  (lower one). Open (closed) symbols represent $L=4$ ($8$). (b) and (c) Spin and lattice-distortion structure factors $F_S({\bf q})$ and $F_L({\bf q})$ in the ordered state in the $(h,h,l)$ plane at $T/J=0.24$ for $L=8$. (d) Spin and lattice-distortion snapshots at $T/J=0.02$. Red (blue) arrows represent up (down) spins and green arrows represent local lattice distortions ${\bf u}^\ast_i$. (e) An extended view of the snapshot of ${\bf u}^\ast_i$ on an $xy$ plane of the lattice, where green arrows represent the $xy$-component with the sign of the $z$-component given together. Gray (orange) colored boxes correspond to gray (yellow) colored tetrahedra in (d). \label{fig:strong}}
\end{figure}

\begin{figure}[t]
\includegraphics[width=\columnwidth]{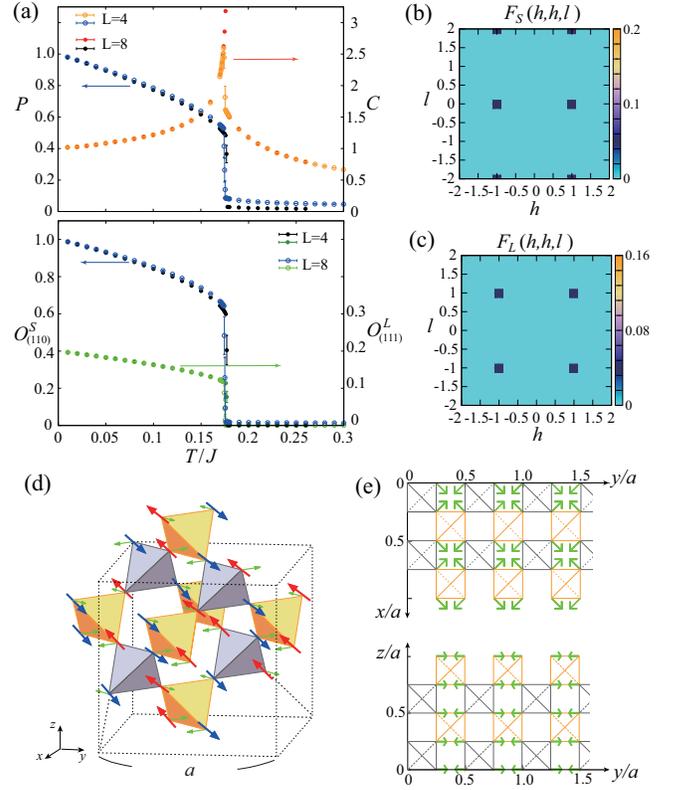}
\caption{MC results for $b=0.2$ obtained in warming runs. Notations are the same as those in Fig.\ref{fig:strong}, while $O^S_{(110)}$ and $O^L_{(111)}$ are defined in the main text. The structure factors [(b) and (c)] and the snapshots [(d) and (e)] are obtained at $T/J=0.12$ and at $T/J=0.01$, respectively. In (e), $xy$ and $yz$ components of the lattice-distortion vectors ${\bf u}^\ast_i$ on $xy$ and $yz$ planes of the lattice are shown.  \label{fig:weak}}
\end{figure}

 Our results are summarized in Fig.\ref{fig:b-T}. With varying the strength of the SLC, $b$, two different types of collinear magnetic phases appear, each accompanied by the commensurate local lattice distortions. For stronger SLC (larger $b$), the ordered phase is cubic symmetric, and the associated spin structure factor $F_{\rm S}({\bf q})$ exhibits multiple Bragg peaks at the $(\frac{1}{2},\frac{1}{2},\frac{1}{2} )$ family, while the lattice-distortion structure factor $F_{\rm L}({\bf q})$ shows Bragg peaks at the cubic-symmetric combination of the $(1,1,0)$ family: see Fig.\ref{fig:strong}. For weaker SLC (smaller $b$), by contrast, the ordered phase is tetragonal symmetric with the spin $F_{\rm S}({\bf q})$ exhibiting multiple Bragg peaks at the cubic-symmetry-broken combination of the $(1,1,0)$ family and the lattice $F_{\rm L}({\bf q})$ showing peaks at the $(1,1,1)$ family: see Fig.\ref{fig:weak}. Both ordered states differ from the one considered in the bond-phonon model \cite{Bond_Penc_04, Bond_Motome_06, Bond_Shannon_10} which is characterized by $(1,1,1)$-type magnetic Bragg patterns. %without the local lattice distortion. 

% Our results are summarized in Fig.\ref{fig:b-T}. With varying the strength of the SLC, $b$, two different types of collinear magnetic phases appear, each accompanied with the commensurate local lattice distortions. For stronger SLC (larger $b$), the ordered phase is cubic symmetric, consisting of $\uparrow \uparrow \downarrow \downarrow$ spin chains running along all six [110] directions. The associated spin structure factor $F_{\rm S}({\bf q})$ exhibits multiple Bragg peaks at the $(\frac{1}{2},\frac{1}{2},\frac{1}{2} )$ family, while the lattice-distortion structure factor $F_{\rm L}({\bf q})$ the ones at the cubic-symmetric combination of the $(1,1,0)$ family: see Fig.\ref{fig:strong}. For weaker SLC (smaller $b$), by contrast, the orfered phase is tetragonal symmetric, consisting of $\uparrow \downarrow \uparrow \downarrow$ spin chains running along two [110] directions and $\uparrow \uparrow \downarrow \downarrow$ ones running along the rest four directions. The associated spin $F_{\rm S}({\bf q})$ exhibits multiple Bragg peaks at the cubic-symmetry-broken combination of the $(1,1,0)$ family. The lattice $F_{\rm L}({\bf q})$ exhibits those at the $(1,1,1)$ family, while the rea-space lattice-distortion pattern is non-cubic (tetragonal): see Fig.\ref{fig:strong}.

 We first derive our model Hamiltonian describing the SLC. In the site phonon model, the displacement vector ${\bf u}_i$ at each site $i$ from its regular position ${\bf r}^0_i$ on the pyrochlore lattice is assumed to be independent of the ones at the neighboring sites (inset of Fig.\ref{fig:b-T}). Although in reality neighboring ${\bf u}_i$'s should be correlated to each other in the form of dispersive phonon modes, we use here the site-phonon model because it is the simplest and minimum model describing phonon-mediated spin interactions. %Of course, if the most relevant phonons in the pyrochlore antiferromagnets are the optical ones rather than the acoustic mode, site phonons would provide a fairly good modelling. 
In the site-phonon picture, an appropriate minimum spin-lattice-coupled model might be
\begin{equation}\label{eq:original_H}
{\cal H} = \sum_{\langle i,j \rangle } J_{\rm ex}\big(|{\bf r}^0_{ij} + {\bf u}_i-{\bf u}_j|\big){\bf S}_i \cdot {\bf S}_j + \frac{c}{2}\sum_i |{\bf u}_i|^2,
\end{equation}
where ${\bf S}_i$ is the classical Heisenberg spin at the site $i$, ${\bf r}^0_{ij} \equiv {\bf r}^0_i-{\bf r}^0_j$, %the distance between the sites $i$ and $j$ in the undistorted setting
$c$ an elastic constant, $J_{\rm ex}$ the exchange interaction %between the sites $i$ and $j$
which is assumed to depend only on the distance between the two spins, and the summation $\langle i,j \rangle$ is taken over all NN pairs. By expanding $J_{\rm ex}$ with respect to ${\bf u}_i/|{\bf r}^0_i|$, i.e., $J_{\rm ex}\big(|{\bf r}^0_{ij} + {\bf u}_i-{\bf u}_j| \big) \simeq J_{\rm ex}\big(|{\bf r}^0_{ij}|\big) + \frac{d J_{\rm ex}}{dr}\Big|_{r=|{\bf r}^0_{ij}|} \, {\bf e}_{ij} \cdot ({\bf u}_i-{\bf u}_j )$ with ${\bf e}_{ij} \equiv {\bf r}^0_{ij}/|{\bf r}^0_{ij}|$, one finds
\begin{eqnarray}\label{eq:distortion}
{\cal H} &\simeq&  J \sum_{\langle i,j \rangle }{\bf S}_i \cdot {\bf S}_j + \frac{c}{2}\sum_i |{\bf u}_i-{\bf u}^\ast_i|^2 - \frac{c}{2}\sum_i |{\bf u}^\ast_i|^2, \nonumber\\
{\bf u}^\ast_i &=& \sqrt{\frac{J \, b}{c}}\sum_{j \in N(i) } {\bf e}_{ij} \, \big( {\bf S}_i \cdot {\bf S}_j \big), 
\end{eqnarray}
where $J\equiv J_{\rm ex}\big(|{\bf r}^0_{ij}|\big)$ and $N(i)$ denotes all the NN sites of $i$ \cite{Site_Jia_05, Site_Bergman_06, Site_Wang_08}. The dimensionless parameter $b= \frac{1}{cJ}\big[ \frac{d J_{\rm ex}}{dr}\big|_{r=|{\bf r}^0_{ij}|} \big]^2$ measures the strength of the SLC. We take $J>0$ and $d J_{\rm ex}/dr < 0$ so that $b>0$. 
Integrating out the ${\bf u}_i$, which is equivalent to minimizing the Hamiltonian with respect to ${\bf u}_i$, we obtain an effective spin Hamiltonian ${\cal H}_{\rm eff}={\cal H}_0+{\cal H}_{\rm SL}$,
\begin{equation}\label{eq:Hamiltonian}
{\cal H}_0 = J \, \sum_{\langle i,j \rangle_s } {\bf S}_i \cdot {\bf S}_j, \quad {\cal H}_{\rm SL}= -\frac{c}{2}\sum_i |{\bf u}^\ast_i|^2 .
\end{equation}
Now, the physical meaning of ${\bf u}^\ast_i$ is clear: it is the optimal local lattice distortion corresponding to the most probable ${\bf u}_i$-value. 
%By contrast, in the bond-phonon model, $u_{ij}={\bf e}_{ij} \cdot ({\bf u}_i-{\bf u}_j)$ is a variable which will be integrated out with the elastic energy $(c/2)\sum_{\langle i,j \rangle} u_{ij}^2$ instead of $(c/2)\sum_i |{\bf u}_i|^2$ in Eq.(\ref{eq:original_H}). In this case, one can in principle determine a distortion not for each {\it site} but for each {\it bond}.
%
%\begin{equation}\label{eq:distortion}
%{\bf u}^\ast_i = \sqrt{\frac{J \, b}{c}}\sum_{j \in N(i) } {\bf e}_{ij} \, \big( {\bf S}_i \cdot {\bf S}_j \big) ,
%\end{equation}
%

 The SLC-term ${\cal H}_{\rm SL}$ can be rewritten into the form
\begin{eqnarray}\label{eq:Hamiltonian_SL}
{\cal H}_{\rm SL} &=& - J \, b \, \sum_{\langle i,j \rangle } \big( {\bf S}_i \cdot {\bf S}_j \big)^2 \nonumber\\
&-& \frac{J \, b}{2}\sum_i \sum_{j\neq k \in N(i) }  {\bf e}_{ij} \cdot {\bf e}_{ik} \, \big( {\bf S}_i \cdot {\bf S}_j \big)\big( {\bf S}_i \cdot {\bf S}_k \big) ,
\end{eqnarray}
where all terms are quartic in ${\bf S}_i$. The first term, whose coefficient is always negative irrespective of the sign of $d J_{\rm ex}/dr$, favors collinear spin states and tends to induce the spin nematic order. The second term includes inter-tetrahedral interactions, namely, effective further neighbor interactions. We emphasize that only the first term exists in the bond-phonon model, so that the ordering characteristics of the present site phonon model would be borne by the second term.

 Since the collinear states are preferred due to the first term of Eq.(\ref{eq:Hamiltonian_SL}), one may assume that the relevant spin states are collinear with a common axis in the spin space, replacing the Heisenberg spin ${\bf S}_i$ with the Ising spin $\sigma_i$. Then, the inter-tetrahedral interactions read $J^{\rm eff}_2\sum_{\langle \langle j,k \rangle \rangle} \sigma_j\sigma_k + J^{\rm eff}_3\sum_{\langle \langle \langle j,k \rangle \rangle \rangle} \sigma_j\sigma_k$ with $J^{\rm eff}_2=J \, b/2$ and $J^{\rm eff}_3=J \, b$ being the effective second- and third-neighbor interactions, respectively. Note that both $J^{\rm eff}_2$ and $J^{\rm eff}_3$ are antiferromagnetic with $J^{\rm eff}_2 < J^{\rm eff}_3$. Due to the strong antiferromagnetic $J^{\rm eff}_3$, three neighboring Ising spins on a straight line avoid to take the up-up-up nor down-down-down configuration. Such a local constraint for the collinear state is called the ``bending rule'' \cite{Site_Bergman_06}. The question then is whether or not these inter-tetrahedral interactions drive the spin LRO for classical Heisenberg spins, and if so, what type. 
% Remember that, in the bond phonon model, only the first term of Eq.(\ref{eq:Hamiltonian_SL}) appears, leading to the spin nematic phase without the standard magnetic LRO unless additional further neighbor interactions are incorporated \cite{Bond_Shannon_10}.

 We investigate the ordering of the model by means of Monte Carlo (MC) simulations. In our simulations, $10^6$ Metropolis sweeps are performed at each temperature with periodic boundary conditions, where the first half is discarded for thermalization. A single spin flip at each site consists of the conventional Metropolis update followed by an over-relaxation update. %of an angle $\pi$. 
The statistical average is taken over $8-10$ independent runs. Total number of spins $N$ is $N=16 L^3$ with $L=4$ and $8$. The results shown below are obtained in the warming runs. Relatively large hysteresis is observed between the warming and the cooling runs due to the first-order nature of the transition (see below) \cite{hysteresis}. 

 Various physical quantities are computed, including the spin-collinearity parameter $P$ %measuring the local spin collinearlity, 
and the structure factors associated with the spin and the lattice distortion, $F_{\rm S}({\bf q})$ and $F_{\rm L}({\bf q})$, each defined by
\begin{eqnarray}\label{eq:phys_quant}
P^2 &=& \frac{3}{2} \Big\langle \frac{1}{N^2}\sum_{i,j} \big( {\bf S}_i\cdot{\bf S}_j\big)^2 - \frac{1}{3} \Big\rangle, \nonumber\\
F_{\rm S}({\bf q}) &=& \Big\langle \Big| \frac{1}{N} \sum_i  {\bf S}_i \, e^{i \frac{2\pi}{a}{\bf q}\cdot{\bf r}_i}\Big|^2\Big\rangle, \nonumber \\
F_{\rm L}({\bf q}) &=& \Big\langle \Big| \frac{1}{N} \sum_i  {\bf u}^\ast_i \, e^{i \frac{2\pi}{a} {\bf q}\cdot{\bf r}_i}\Big|^2\Big\rangle,
\end{eqnarray}
where $\langle O \rangle$ denotes the thermal average of a physical quantity $O$, and ${\bf q}$ is a wave number in units of $2\pi/a$. The lattice distortion ${\bf u}^\ast_i$ is given in units of $\sqrt{J/c}$.

 Simulation results for $b=0.35$, which corresponds to the strong SLC regime, are shown in Fig.\ref{fig:strong}. In Fig.\ref{fig:strong}(a), a sharp peak in the specific heat $C$  accompanied by a discontinuous jump in the spin collinearity $P$ suggests the occurrence of a first-order transition into the ordered state with the long-range spin collinearity. The spin and the lattice-distortion structure factors in the ordered phase are shown in Figs.\ref{fig:strong}(b) and (c), respectively. The spin $F_{\rm S}({\bf q})$ exhibits Bragg peaks of equal heights at all $(\pm \frac{1}{2}, \pm \frac{1}{2}, \pm \frac{1}{2})$ and $(\pm \frac{3}{2}, \pm \frac{3}{2}, \pm \frac{3}{2})$ points, indicating the magnetic LRO keeping the cubic symmetry. Since the local lattice distortion ${\bf u}_i^\ast$ is directly connected to the spin via Eq.(\ref{eq:distortion}), ${\bf u}_i^\ast$ also exhibits a LRO of cubic symmetry as is evidenced by multiple Bragg peaks of equal-heights in the lattice $F_{\rm L}({\bf q})$ observed at all $(\pm 1,\pm 1,0)$, $(\pm 1,0, \pm 1)$ and $(0, \pm 1,\pm 1)$ points \cite{110-peak}. 

 The spin and the lattice-distortion real-space configurations taken from a snapshot of our simulations are shown in Figs.\ref{fig:strong}(d) and (e). As can be seen in (d), the $\uparrow \uparrow \downarrow \downarrow$ spin chains run along all six [110] directions, keeping the cubic symmetry of the lattice. 
In units of tetrahedron, this collinear order consists of a periodic arrangement of six two-up two-down, four three-up one-down, four one-up three-down, one all-up, and one all-down tetrahedra. %out of sixteen tetrahedra in a cubic unit cell, six (two-up two-down), four (three-up one-down), four (one-up three-down), one (all-up), and one (all-down) tetrahedra occur. 
The corresponding lattice-distortion ${\bf u}^\ast_i$ pattern is shown in Fig.\ref{fig:strong}(e) in the form of the $xy$-plane projection. The lattice distortion is also cubic symmetric, corresponding $xz$ and $yz$ projections of ${\bf u}^\ast_i$ (not shown here) being similar to the $xy$ one.

 While the $(\frac{1}{2},\frac{1}{2},\frac{1}{2})$-type magnetic order looks scarce in the pure Heisenberg spin model, it was reported in the spin-ice Kondo-lattice model \cite{Ice_Ishizuka_11} where our `up' and `down' spins correspond to `in' and `out' spins. A common feature of the two models might be the existence of a relatively strong third-neighbor antiferromagnetic interaction. We note in passing that the same $(\frac{1}{2},\frac{1}{2},\frac{1}{2})$ magnetic Bragg patterns were observed in the spin-ice model with the long-range RKKY interaction \cite{Ice_Ikeda_08}, though the spin structure is not completely the same.

 Next, we discuss the weak SLC regime of smaller $b$. Figure \ref{fig:weak} shows our simulation results for $b=0.2$. One can see that the system exhibits a first-order transition into the collinearly-ordered spin state with the magnetic Bragg peaks at $(\pm 1,\pm 1,0)$ (see $F_{\rm S}({\bf q})$ of Fig.\ref{fig:weak}(b)). Although Fig.\ref{fig:weak}(b) looks similar to the lattice $F_{\rm L}({\bf q})$ in the strong SLC regime shown in Fig.\ref{fig:strong}(c), the state here is {\it not} cubic symmetric, spontaneously breaking the cubic symmetry. The state is tetragonal symmetric in that, among three equivalent $(0, 1, 1)$, $(1, 0, 1)$, and $(1, 1, 0)$ states, only one is selected \cite{110-peak}. In the lower panel of Fig.\ref{fig:weak}(a), we show the temperature dependence of the average Bragg intensity defined by $O^{\rm S}_{(110)} = 2\sum_{n,m =\pm 1} \big[ F_{\rm S}(0,n,m) + F_{\rm S}(n,0,m) + F_{\rm S}(n,m,0) \big]$. The associated real-space spin configuration is shown in  Fig.\ref{fig:weak}(d). The $\uparrow \downarrow \uparrow \downarrow$ spin chains run along the $[110]$ and $[1 \overline{1}0]$ directions, while the $\uparrow \uparrow \downarrow \downarrow$ chains run along the rest four directions. We note that this collinear order consists only of two-up two-down tetrahedra and is the same as that of Ref.\cite{SLC_Tchernyshyov_prl_02, SLC_Tchernyshyov_prb_02, SLC_Chern_06} obtained by the phenomenological Landau analysis.

 The corresponding lattice $F_{\rm L}({\bf q})$ is shown in Fig.\ref{fig:weak}(c), where the Bragg peaks are observed at all $(\pm 1, \pm 1, \pm 1)$ points. The average Bragg intensity defined by $O^{\rm L}_{(111)} = (1/8) \sum_{h,k,l =\pm 1} F_{\rm L}(h,k,l)$ is shown in the lower panel of  Fig.\ref{fig:weak}(a). Although the lattice distortion expected from the $F_{\rm L}({\bf q})$ peaks might look like cubic symmetric, this is actually not the case: the tetragonal symmetry of the spin order results in a two-dimensional lattice distortion. %pattern of the full lattice-distortion vectors ${\bf u}^\ast_i$. 
Namely, when the $(1, 1, 0)$-type spin order is selected, for example, ${\bf u}^\ast_i$ lie only in the $xy$ plane with $u^\ast_{i, z}=0$ as shown in (e). Indeed, the results shown in Figs.\ref{fig:weak}(b)-(e) are the ones associated with the $(1, 1, 0)$ spin ordered state. 
%In fact, we performed our warming-run simulations from spin initial conditions corresponding to one of three possible cubic-symmetry-broken state.

\begin{figure}[t]
\includegraphics[width=\columnwidth]{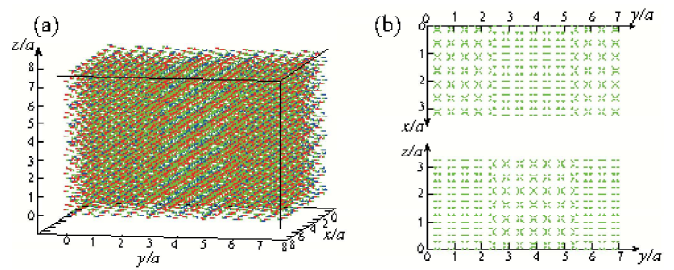}
\caption{(a) A MC snapshot of the spin and the lattice-distortion patterns.  Notations of the symbols are the same as those in Figs.\ref{fig:strong} and \ref{fig:weak}. (b) Snapshots of the lattice distortion ${\bf u}^\ast_i$ projected onto the $xy$ and the $yz$ planes. \label{fig:domain}}
\end{figure}

 In cooling runs from a high temperature \cite{hysteresis}, one often encounters domain states consisting of different $(1,1,0)$ domains, an example of which is shown in Fig.\ref{fig:domain}(a). The $(0,1,1)$ ordered state is intercalated as a domain in the middle of the $(1,1,0)$ state, with a domain wall lying in the $xz$ plane. %Figure \ref{fig:domain}(a) shows an example of a MC snapshot with such a domain structure, where the $(0,1,1)$ ordered state is intercalated by the $(1,1,0)$ domain in the middle, with a domain wall lying in the $(xz)$ plane. 
The associated lattice-distortion pattern is shown in Fig.\ref{fig:domain}(b). 
Although the domain formation usually costs energy, a planar and flat domain wall spanning over the whole system like the one shown in Fig.\ref{fig:domain} costs no energy in the present model. This is because the bending rule is satisfied even at the domain wall, as long as the planar domain wall extends over the whole system. Hence, domain walls, once generated, stay quite stable in the present model, and are likely to be so also in relevant experimental systems.
% Closer examination reveals that the phase boundary between the cubic $(\frac{1}{2},\frac{1}{2},\frac{1}{2})$ and the tetragonal $(1,1,0)$ spin ordered phases around $b_c\sim 0.25$ exhibits a nontrivial `reentrant' behavior. If the temperature is lowered from the high-temperature phase at $b=*$, for example, the system exhibits successive paramagnetic $\rightarrow$ tetragonal $\rightarrow$ cubic *** transitions.
 
 We finally discuss possible experimental relevance of our results. In neutron measurements on various chromium spinels ACr$_2$O$_4$ (A=Zn, Cd, Hg) and LiInCr$_4$O$_4$ \cite{ZnCrO_Lee_08, HgCrO_Matsuda_07, CdCrO_Chung_05, BrPyro_Nilsen_15}, rich and complex Bragg-peak patterns have been observed, but they basically involve $(1,1,0)$ reflections as observed in the weak SLC regime of the present model (also suggested in Ref.\cite{SLC_Tchernyshyov_prl_02, SLC_Tchernyshyov_prb_02, SLC_Chern_06}). 
%Neutron measurements performed for various chromium spinels ACr$_2$O$_4$ (A=Zn, Cd, Hg) and LiInCr$_4$O$_4$ \cite{ZnCrO_Lee_08, HgCrO_Matsuda_07, CdCrO_Chung_05, BrPyro_Nilsen_15} yield Bragg-peak patterns rich and complex, but basically involving $(1,1,0)$ reflections as observed in the weak SLC regime of the present model (also suggested in Ref.\cite{SLC_Tchernyshyov_prl_02, SLC_Tchernyshyov_prb_02, SLC_Chern_06}). 
Bragg reflections at $(\frac{1}{2},\frac{1}{2},\frac{1}{2})$ observed in the strong SLC regime of the present model have also been reported as a magnetic domain in ZnCr$_2$O$_4$ \cite{ZnCrO_Lee_08}, although the experimentally proposed spin structure appears to be different from our present one. These results suggest that the SLC originating from site phonons may be relevant to the magnetic ordering in chromium spinel oxides.

 In our analysis, no {\it net} lattice distortion is considered, in contrast to the experimental cubic-tetragonal or cubic-orthorhombic structural transitions observed in chromium spinels. However, since the existence of non-uniform lattice distortions in addition to the uniform net ones has been suggested even experimentally \cite{ZnCrO_Lee_00, ZnCrO_Ueda_03}, we believe that the present site phonon model taking account of the local lattice distortion captures the essential part of the ordering mechanism. In principle, the uniform net deformation could be considered by incorporating dispersive phonon modes beyond the site-phonon model, but this issue, including the associated spin structure, is beyond the scope of the present paper.
 
To our knowledge, there have been no detailed diffraction studies to detect possible {\it local\/} lattice distortions of chromium spinels. It would then be interesting to perform appropriate neutron or X-ray diffraction measurements to detect local lattice distortions. %Also, optical spectroscopic measurements which are sensitive to the crystal symmetry might be useful in detecting the signature of such local distortions \cite{ZnCrO_Miyata_prl_11, ZnCrO_Miyata_jpsj_12}.

%%%%%%%%%%%%%%%%    
%We note that the $(1,1,0)$ collinear spin structure has been also reported in AV$_2$O$_4$ (A=Zn, Cd, Mg) \cite{Zn-CrVO_Zhang_06, MgVO_Niziol_73}. Although the present model cannot be directly applied to AV$_2$O$_4$ because structural and Neel transitions do not occur simultaneously in these materials, it may give a clue to understand the origin of the Neel state. 
%%%%%%%%%%%%%%%%
  
 To conclude, we have shown that the spin-lattice coupling originating from site phonons leads to two types of commensurate collinear magnetic orders accompanied by the local lattice distortions, both of which are likely to be relevant to the ordered states of chromium spinels. This suggests that the site phonons could play a key role in a class of frustrated magnets, giving rise to rich spin-lattice-coupled orderings.

 The authors thank T. Okubo, Y. Motome, Y. Okamoto and K. Tomiyasu for useful discussion and comments. They are thankful to ISSP, the University of Tokyo for providing us with CPU time. This study is supported by a Grant-in-Aid for Scientific Research No. 25247064.


\begin{thebibliography}{50}
\bibitem{hysteresis} On cooling, the first order transition temperatures at $b=0.2$ and $b=0.35$ are lower than the corresponding warming results by 10\% and 20\%, respectively, but the low-temperature spin structures themselves are basically unchanged.
\end{thebibliography}
\end{document}